# From Brittle to Ductile: A Structure Dependent Ductility of Diamond Nanothread

Haifei Zhan[1,2,3], Gang Zhang[2,*], Vincent BC Tan[3], Yuan Cheng[2], John M. Bell[1], Yong-Wei Zhang[2], and Yuantong Gu[1,*]

[1]*School of Chemistry, Physics and Mechanical Engineering, Queensland University of Technology (QUT), Brisbane QLD 4001, Australia*
[2]*Institute of High Performance Computing, Agency for Science, Technology and Research, 1 Fusionopolis Way, Singapore 138632*
[3]*Department of Mechanical Engineering, National University of Singapore, 9 Engineering Drive 1, Singapore 117576*

**Abstract:** As a potential building block for the next generation of devices/multifunctional materials that are spreading almost every technology sector, one-dimensional (1D) carbon nanomaterial has received intensive research interests. Recently, a new ultra-thin diamond nanothread (DNT) has joined this palette, which is a 1D structure with poly-benzene sections connected by Stone-Wales (SW) transformation defects. Using large-scale molecular dynamics simulations, we found that this $sp^3$ bonded DNT can transit from a brittle to a ductile characteristic by varying the length of the poly-benzene sections, suggesting that DNT possesses entirely different mechanical responses than other 1D carbon allotropies. Analogously, the SW defects behave like a grain boundary that interrupts the consistency of the poly-benzene sections. For a DNT with a fixed length, the yield strength fluctuates in the vicinity of a certain value and is independent of the "grain size". On the other hand, both yield strength and yield strain show a clear dependence on the total length of DNT, which is due to the fact that the failure of the DNT is dominated by the SW defects. Its highly tunable ductility together with its ultra-light density and high Young's modulus makes diamond nanothread ideal for creation of extremely strong three-dimensional nano-architectures.

**Introduction**

Low dimensional crystalline carbon nanomaterials/nanostructures, such as $sp^3$ diamond nanowire,[1] $sp^2$ carbon nanotube (CNT),[2] and $sp^2$ graphene,[3] have been witnessed with intensive interest from both scientific and engineering communities due to their superior mechanical, versatile chemical, fascinating physical and electrical properties, which have enabled them as building blocks for a vast range of usages in the next generation of devices[4, 5] or multifunctional materials[6] (from 1D nano-fibers/yarns[7, 8] to 2D nanomesh,[9] and 3D porous structures[10]). Specifically, the $sp^3$ bonded diamond nanowires, which possess unique features, such as negative electron affinity, chemical inertness, good biocompatibility, have received a continuing research focus.[11]

Recently, a new 1D $sp^3$ carbon nanomaterial has been reported, which is synthesized through solid-state reaction of benzene under high-pressure, termed as diamond nanothread (DNT).[12] On the one hand, the DNT is a close-packed $sp^3$-bonded carbon structure, with carbon atoms arranged in a diamond-like tetrahedral motif (see Figure 1), similar to the diamond nanowire/nanorod. On the other hand, the DNT can be regarded as hydrogenated (3,0) CNTs connected with Stone-Wales (SW) transformation defects (see inset of Figure 1).[13] While, unlike CNTs, the existence of SW transformation defects interrupts the central hollow of the structure. Thus, a fundamental understanding of how the mechanical properties of the DNT differ from its counterpart (i.e., CNT and diamond nanowire) is of great interest.

Previous studies have shown that the carbon nanotubes exhibit outstanding mechanical strength. For example, experimentally measured tensile Young's modulus for SWNTs ranges from 320 GPa to 1.47 TPa with the breaking strengths ranging from 13 to 52 GPa, and the breaking strain up to 5.3%.[14] Whereas, a first-principles calculation suggests that the diamond nanowire has a low Young's modulus ranging from ~ 40 to 290 GPa.[15] Unfortunately, most diamond nanostructures are brittle and easy to fail under tensile load, which limits their applications as a building block in the nanoscale. Very recently, a preliminary study[16] has shown that DNT has excellent mechanical properties, namely, a high stiffness of about 850 GPa, and a large bending rigidity of about $5.35 \times 10^{-28}$ N·m$^2$. A few questions arise promptly: what is the ductility of DNT? How does the sample length affect the mechanical properties? How its internal structure affects the mechanical properties? Clearly, answers to these questions are crucial for the technological explorations.[17-20] To this end, in this work, we explore the mechanical characteristics of DNTs through the investigations of the length-dependency and structural-influence on their mechanical properties. From large-scale molecular dynamics simulations, we found that the DNT can transit from a brittle to a ductile behaviour, which is benefited from the ductile characteristic of the constituent SW transformation defects. This phenomenon has not been observed in other diamond nanostructures. It is worth noting that different from the bulk material, the diamond nanothread is an ultra-thin nanomaterial, which simply can't have traditional plastic deformation, e.g., dislocations. Therefore, our discussions regarding the "brittle" or "ductile" characteristic are based on the context of the stress-strain curve.

**Computational Methods**

The mechanical behaviors of DNTs were acquired based on a series of tensile tests performed using large-scale molecular dynamics (MD) simulations. To initiate the simulation, the widely used adaptive intermolecular reactive empirical bond order (AIREBO) potential was employed to describe the C-C and C-H atomic interactions.[21] This potential has been shown to well represent the binding energy and elastic properties of carbon materials. It should be noticed that the AIREBO potential usually suffers from a nonphysical high tensile stress which is originated from the fixed switching function.[22] To overcome this problematic issue, the cut-off distance is usually extended far from the original value 1.7 Å to ~ 1.9 – 2.0 Å.[23-28] For the DNT structure, our calculations suggest that a cut-off distance between 1.94 - 1.95 Å for AIREBO potential would result in a comparable yield strain with that obtained from the reactive force filed (ReaxFF)[29] (see discussions in Supporting Information). Thus, a cut-off distance of 1.945 Å is adopted in all the simulations except further declaration.

The DNT structures were firstly optimized by the conjugate gradient minimization method and then equilibrated using Nosé-Hoover thermostat[30, 31] for 2 ns. Periodic boundary conditions were applied along the length direction during the relaxation process. To limit the influence from the thermal fluctuations, a low temperature of 50 K was adopted. The tensile testing was achieved by applying a constant strain rate (namely, $10^{-7}$ fs$^{-1}$) to the fully relaxed models, while keeping the periodic boundary conditions along the length direction. The simulation was continued until the failure of the DNT. A small time step of 0.5 fs was used for all above calculations with all the MD simulations being performed under the software package LAMMPS.[32]

During the tensile simulation, the commonly used virial stress was calculated, which is defined as[33]

$$\Pi^{\alpha\beta} = \frac{1}{\Omega} \left\{ -\sum_i m_i v_i^\alpha v_i^\beta + \frac{1}{2} \sum_i \sum_{j \neq i} F_{ij}^\alpha r_{ij}^\beta \right\} \qquad (1)$$

Here, $\Omega$ is the volume of the system; $m_i$ and $v_i$ are the mass and velocity of atom $i$; $F_{ij}$ and $r_{ij}$ are the force and distance between atoms $i$ and $j$; and the indices $\alpha$ and $\beta$ represent the Cartesian components. Considering that the DNT's analogue – (3,0) CNT has a diameter of 2.35 Å, smaller than the graphite interlayer distance (namely, 3.52 Å), we adopted a solid cylinder to approximate the DNT's volume. The approximate distance between exterior surface hydrogens (i.e., ~ 0.5 nm) was adopted as the diameter of the cylinder following Roman *et al.*[16] Further, the atomic virial stress was estimated according to Eq. (1) as

$$\pi_i^{\alpha\beta} = \frac{1}{\bar{\omega}_i} \left\{ -m_i v_i^\alpha v_i^\beta + \frac{1}{2} \sum_{j \neq i} F_{ij}^\alpha r_{ij}^\beta \right\} \qquad (2)$$

where $\bar{\omega}_i$ represents the effective volume of atom $i$ and $\Omega = \sum \bar{\omega}_i$. With the obtained overall stress, the yield strain is defined as the strain threshold value where the stress shows abrupt reduction and the structure starts to fail. The corresponding stress is designated as the yield strength. For comparison purpose, the effective Young's modulus of the DNT is extracted from the stress-strain curve using linear regression. Based on the assumption of linear elasticity, the initial linear regime has been selected with the strain up to 3%. Such approach has been widely applied to evaluate the mechanical properties of nanomaterials, and validated by earlier studies.[16, 34-36]

**Results and Discussions**

The diamond nanothread (DNT) models were established based on recent experimental observations and first-principles calculations.[12] As illustrated in Figure 1, the DNT contains two different sections, including the Stone-Wales (SW) transformation defect and poly-benzene rings. Of note, the poly-benzene rings here are equivalent to the hydrogenated (3,0) carbon nanotubes. Also, the SW transformation defect here represents the defective structure resulted from the 90° rotation of a C-C dimer, which is different from the commonly discussed pentagon-heptagon pair in CNTs or graphene. For discussion simplicity, we emphasized on the DNT structures with evenly distributed SW defects, and a DNT unit cell with $n$ poly-benzene rings between two adjacent SW defects is denoted by DNT-$n$. In a recent work based on first principle calculations,[37] several different $sp^3$ nanothreads are predicted theoretically. The structures we focused on are the energetically favorable ones. It is worth mentioning that adding one SW defect to the structure will increase the total system energy by ~12 Kcal/mol.[38] With this consideration, even the DNT with the highest defect density (DNT-2) examined in this work is energetically favorable structure.

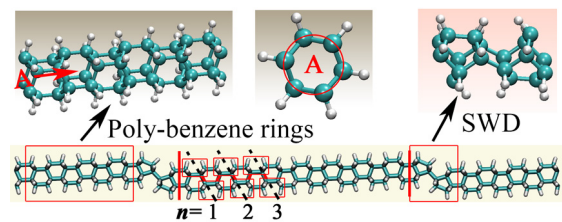

**Figure 1** The atomic configurations of a segment of DNT, insets show the structural representation of the poly-benzene rings and the Stone-Wales transformation defect (SWD).

*Length dependency*

Firstly, we investigate the length dependency of the DNT's tensile properties by examining three groups of samples that are constructed by repeating DNT-8, DNT-14 and DNT-20 units, respectively. Figure 2a and 2b show a clear correlation between the yield strain/strength and the DNT sample length. In detail, the yield strain exhibits a relatively sharp reduction in the region when $L$ is smaller than ~ 30 nm, and then saturates to a certain value. Such changing profile is found uniformly in all three groups. From Figure 2a, the DNT comprised of shorter poly-benzene units tends to saturate to a higher yield strain. For example, with $L > 40$ nm, the averaged yield strain for DNT-8 is 9.0 ± 0.3 %, which is about 15% larger than that of the DNT-14 (about 7.8 ± 0.3 %).

In comparison, the yield strength shows a similar changing pattern as presented in Figure 2b. For instance, the yield strength experienced more than 25% reduction (from ~ 75 GPa to ~ 56 GPa) for the DNT-14 when the sample length increases from

~ 13 nm to 26 nm. Afterward, it fluctuates around 56 GPa. Unlike the yield strain, the yield strength for all considered DNTs saturates to a similar value (around 56 GPa), and exhibits a relation irrelevant with the constituent units for the investigated length scope (from ~ 13 – 92 nm). Recall the morphology of the DNT (Figure 1), the SW defect is analogue to the grain boundary and the constituent unit (poly-benzene) length is equivalent to the grain size. In other words, the yield strength of the DNT remains constant with decreasing grain size (i.e., the poly-benzene length). Further evidence for such relationship was found when we assessed the structural influence on the mechanical properties of the DNT as discussed in the following section. It is worth noting that the strain rate would also influence the yield strength/strain. To acquire such impacts, a 24 nm DNT-17 was considered. Tested strain rate includes $1\times10^{-6}$, $5\times10^{-7}$, $1\times10^{-7}$, $5\times10^{-8}$, and $1\times10^{-8}$ fs$^{-1}$, as shown in the Supporting Information (Figure S4). We find that the DNT exhibits different yield strength/strain at different strain rates, and higher strain rate leads to larger yield strength/strain. Specifically, the yield strength/strain shows slight difference for extremely low strain rate (less than $1\times10^{-7}$). In-detail analysis show that although the DNT exhibits different yield strength/strain, the deformation mechanisms are the same, i.e., stress concentration at the locations of SW transformation defect and the failure originated from these locations. These results signify that the strain rate ($10^{-7}$ fs$^{-1}$) considered in this work exerts ignorable impacts on the tensile behaviour of the studied DNT, and it is suitable for the investigation purpose.

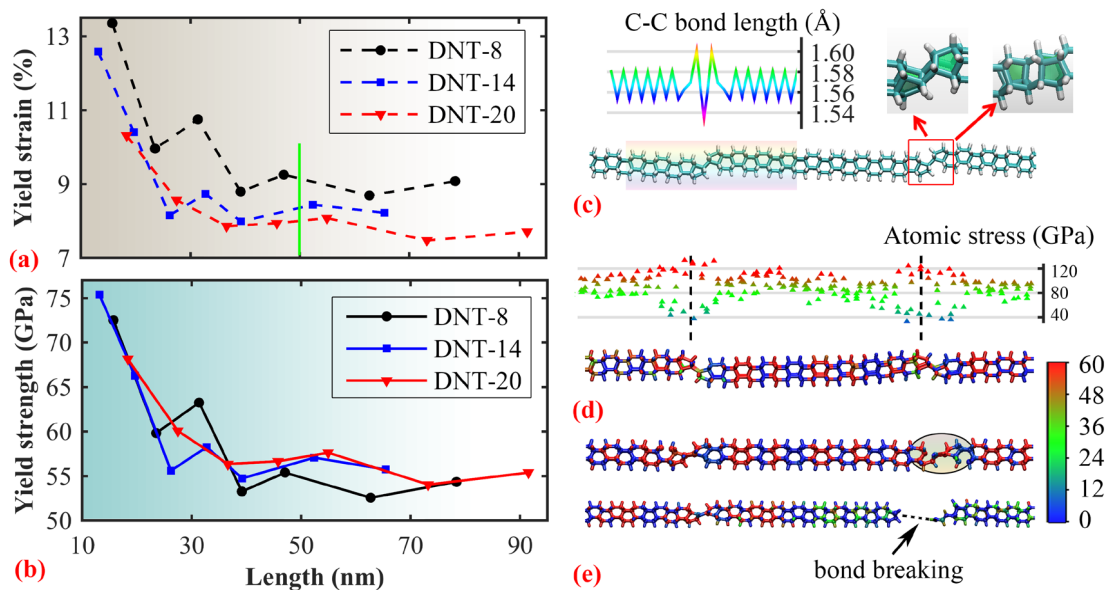

**Figure 2** (a) Estimated yield strain; and (b) yield strength of the DNT constructed by DNT-8, DNT-14 and DNT-20 units. (c) The C-C bond length distribution in the DNT, insets show the pentagons and hexagon that realized the SW defect; (d) The virial atomic stress distribution along the length direction at the strain of 4.6%, which clearly shows the stress concentration at the SW defect region (only carbon atom stress is presented); (e) The bond breaking configuration at the pentagon (upper, strain of 11.2%), which eventually initiates the failure of the DNT from the SW defect (lower, strain of 13.4%).

To explain the length dependency of the yield strain/strength, we inspect the deformation processes of the DNT. Figure 2d shows the virial atomic stress distribution along the length direction at the strain of 4.6%. It is clear that the SW defect regions are subjected to higher stress during tensile deformation, i.e., stress concentration. Such observation is attributed to the initial bond length distribution as plotted in Figure 2c, from which a much longer C-C bond in the pentagons (comprising the SW defect) are observed. Thus, the SW defect regions experience a higher atomic stress after stretching. On the other hand, such stress concentration also dominates the failure mode of the DNT. From Figure 2e, initial bonding breaking is found in the pentagons, which eventually leads to the failure of the whole structure with increasing elongation. Such deformation mode has been observed uniformly from the examined DNTs made from DNT-8, DNT-14 and DNT-20 units. Our simulation results demonstrate that there is a stress concentration around the SW defect region during tensile deformation. Therefore, longer DNTs possess more SW defects, meaning that they have more stress concentration regions and stronger local variances, and thus are easier to fail (i.e., leads to lower yield strain and yield strength). [39]

Surprisingly, although the yield strain/strength exhibits a clear length dependent characteristic, the estimated Young's modulus within each group is barely influenced by the sample length. From Figure 3, the estimated Young's modulus for DNT-8 shows a slight reduction when *L* increases from 15.7 to 78.4 nm (from 831.8 to 799.1 GPa). Similar trend is also observed for other two groups comprised by DNT-14 and DNT-20, with the average Young's modulus as 872.5 ± 1.2 and 898 ± 10.3 GPa, respectively. Evidently, although the total length has insignificant influence on the Young's modulus of the DNTs, an apparent difference exists among the three groups. For instance, the average Young's modulus for DNT-8 is about 11% smaller compared with that of the DNT-20. This reveals a strong dependence of mechanical properties on the structure of DNTs. Such phenomenon can be explained from the perspective of the structural influence as detailed below.

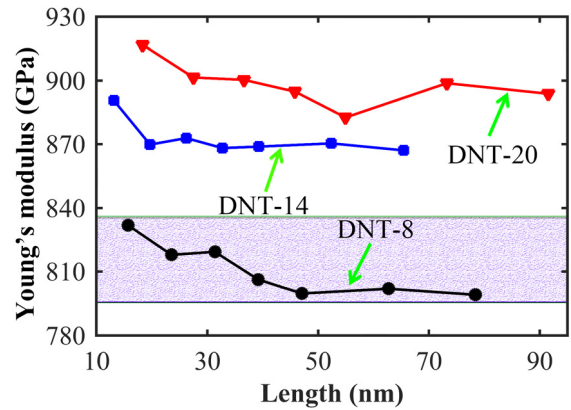

**Figure 3** Comparisons of Young's modulus extracted from different DNTs (constructed from DNT-8, DNT-14 and DNT-20 units) with varying sample length.

### *Structure –mechanical property correlation: from brittle to ductile*

The above discussions have shown that the DNT has a length-dependent mechanical property. Another interesting question that is worth exploring is whether its mechanical behaviors can be tuned through the structural change, say, varying the number of SW defects in a given sample length. To answer this, we constructed DNTs with a fixed length of about 42 nm, and compared the tensile properties of DNTs containing different number of SW defects by changing the length of DNT-*n*. From Figure 4a, it is obvious that the DNT-*n* with longer poly-benzene (larger *n*) exhibits a classical brittle behavior with a monotonically increased stress-strain curve; whereas, the DNTs with short poly-benzene, such as DNT-2, shows a clear hardening process besides the monotonically increased portion. The most interesting feature is that the hardening process has greatly deferred the failure of the DNT. For example, the sample DNT-2 (has 32 SW defects) with shorter poly-benzene has a yield strain nearly twice of its counterpart DNT-48, which is comprised of longer poly-benzene (2 SW defects). More strikingly, the hardening duration is found to extend gradually with the decrease of the constituent poly-benzene length, signifying an evident transition of the DNT from a brittle to a "ductile" behavior. Such observation has been uniformly observed from the three examined groups of the DNT with the length of 24, 31, and 42 nm. Explanations for such novel observation are given later.

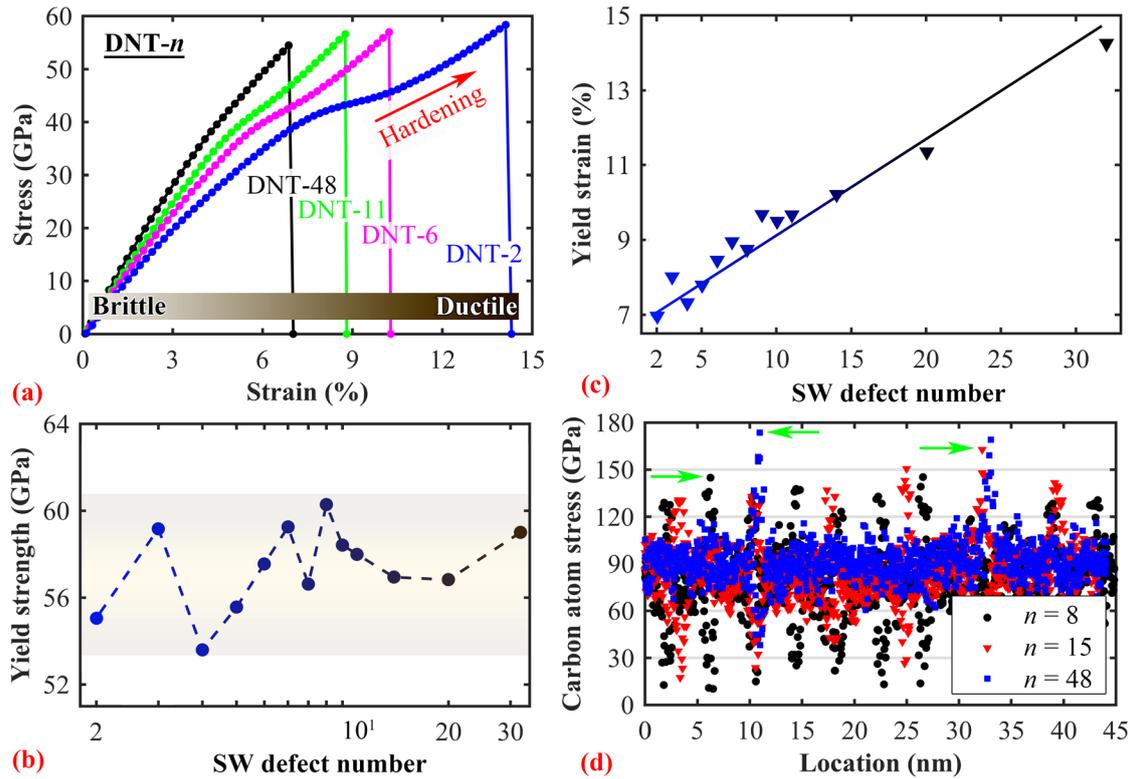

**Figure 4** Numerical results from the DNTs with a uniform sample length of ~ 42 nm: (a) Stress-strain curves from DNTs comprised of by different constituent units; (b) The estimated yield strength, which fluctuates around a certain value, indicating a relation irrelevant with the number of SW defects; (c) The corresponding yield strain *versus* the number of SW defects; (d) Virial atomic stress at the strain of 5% for DNT-8, DNT-15, DNT-48 with 11, 6, and 2 SW defects (only carbon atom stress is presented). The green arrows highlighted the maximum stress in the stress concentration regions.

Despite the transition phenomenon, we found that the estimated yield strength fluctuates in the vicinity of a certain value (Figure 4b). Such phenomenon indicates a relationship irrelevant with the number of SW defects, same as that observed from Figure 2b. The average value for the sample length of 42 nm is 57.4 ± 1.9 GPa. Recall the atomic configurations in Figures 2d and 2e, the underlying mechanism for such relationship is that the failure normally happens at the SW defect region (when the maximum tolerable stress of the SW defect is reached). The fluctuations of the yield strength are originated from the different stress distributions/localization variance at the locations of the SW defects, which are vulnerable to thermal perturbations. As evidenced from our simulations, different stress distribution patterns occur at the locations of SW defects (see Supporting Information). In the current simulation we adopted a low temperature of 50 K, higher temperature is likely to influence the stress concentration and thus lead to different yield strength. A comprehensive investigation of such temperature-dependent mechanical behavior of DNT will be the focus of future study.

Unlike the yield strength, the yield strain exhibits a general increasing relationship with the number of SW defects as illustrated in Figure 4c (the results from sample length of ~ 24 and 31 nm exhibit a similar profile). The increasing trend is not contradicted with the previously observed length-dependent characteristic. Recall Figure 2a, if we consider a same sample length (the vertical green line), the DNT with longer poly-benzene section (i.e., less SW defects) also tends to have a smaller yield strain. Such results are on the one hand benefited from the extended hardening process, and on the other hand, due to the alleviated stress concentration. For a DNT with fixed sample length, adding more SW defects will reduce the maximum stress in the concentration area at the same strain. As evidenced in Figure 4d, the maximum stress in the stress concentration region is on average smaller for the structure with more SW defects. Therefore, it is reasonable to observe an increasing yield strain with the number of SW defects for a given sample length, which on the other hand affirms the increased ductility of the DNT.

*The ductile characteristic of the SW transformation defect*

With above discussions, we then exploit the origins for the novel transition observed in Figure 4a. Specifically, we track the stress-strain relation in a confined region with only poly-benzene rings or the SW defect. To achieve this, we freeze the irrelevant regions and introduce a linear velocity field to stretch the targeted area (see Supporting Information). Also, the cut-off distance was tuned to a large value of 2.0 Å to ensure that the transition phenomenon is not originated from the deficiency of the AIREBO potential. As expected, we find that the confined region with only poly-benzene rings exhibits a classical brittle behavior (curve P-20), which is not affected by increasing the region length/scope (curve P-182 in Figure 5a), signifying a brittle characteristic of the poly-benzene sections. However, for the SW defect, an extra hardening process is observed (black stress-strain curve 5 in Figure 5a), which endows it with a yield strain approaching 25%, more than twice of that extracted from the confined region with only poly-benzene rings. Such results imply the ductile characteristic of the SW defect region, which is resulted from the initial bond breaking at the pentagon carbon rings. In-depth analyses show that only the bond-stretching process was involved during the first stress increase portion (A-B in Figure 5a). When the strain approaches 15%, two C-C bonds of the pentagons are found to break (see inset of Figure 5a), which initiates the hardening process (B-C portion, refer to Supporting Information for more details). The breaking of these bonds endows the SW defect region with greatly extended deformability before fracture, which we term as a ductile characteristic. Due to the different structures, we should emphasis that the ductile characteristic discussed herein is different from the movement of 5-7 structural defects as discussed in CNT.[40] Moreover, a fixed temperature of 50 K was used in our simulations. The brittle-to-ductile transitions are typically temperature dependent. Based on the unique structural property of DNT, the similar brittle-to-ductile transition is reasonably expected at room temperature. However, the comprehensive temperature impacts on the brittle-to-ductile transition deserve further studies.

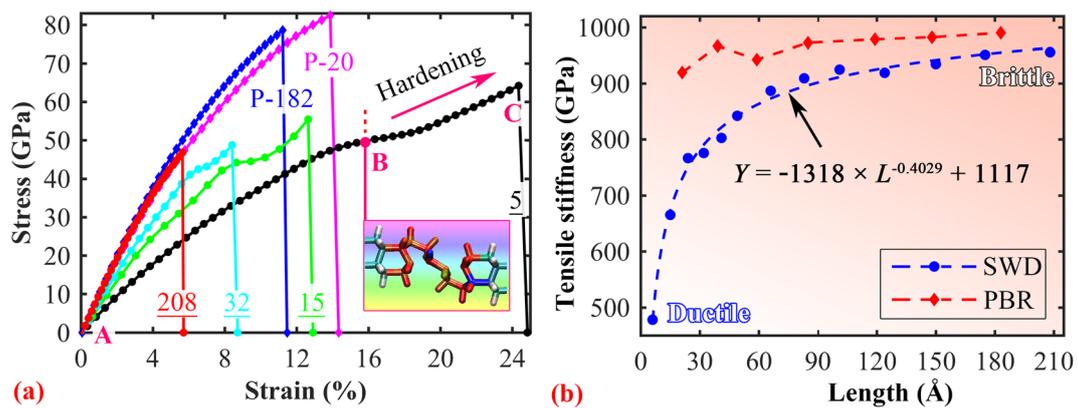

**Figure 5** (a) The stress-strain curves obtained from different confined regions with poly-benzene rings and SW defect. P-20 and P-182 denote the two confined regions contain only poly-benzene rings with a length of about 20 and 182 Å, respectively. The other underline numbers represent the approximate length of the confined region with one SW transformation defect. Inset shows the atomic configurations of the SW defect at the strain of 16.1%. (b) The tensile stiffness as a function of the length of the confined regions. SWD and PBR represent the confined region with SW transformation defect and poly-benzene ring, respectively. Clearly, the tensile stiffness for the purely poly-benzene rings is independent of the constrained region length.

Therefore, for a fixed sample length with a small number of SW defects, the DNT's mechanical behavior is dominated by the poly-benzene sections, i.e., exhibiting a brittle behavior. *Vice versa*, the DNT yields to a ductile behavior when more SW defects are introduced. In other words, the ductility of DNT can be controlled by altering the number of SW defects. As illustrated in Figure 5a, by continuing adding poly-benzene rings to the two ends of the confined SW transformation defect region, its ductile characteristic is gradually suppressed, and eventually the DNT exhibits a brittle behavior (red stress-strain curve 208 in Figure 5a). Accompanying with this transition process, the effective Young's modulus which is a placeholder for the tensile stiffness of the confined regions (blue curve) firstly experiences an exponential increase and then converges to the value extracted from the purely poly-benzene rings (red curve). As plotted in Figure 5b, the tensile stiffness of the SW defect region (blue curve) increases almost 100% (from ~ 480 to 900 GPa), showing the transition from a ductile characteristic to the brittle characteristic. These results suggest a highly tailorable mechanical property of the DNT endowed by its intriguing structure. In this respect, experimental efforts are expected to find a way to tune the density of SW transformation defects or the number of poly-benzene rings (grain size), perhaps via the control of the benzene distributions, molecular weights or the reaction pressure.[12]

Of interest, we also compare the mechanical behaviour of the DNT with the serial spring model proposed by Roman et al.[16] Basically, the DNT is simplified as a system connected by two types of springs, with one representing the poly-benzene section (PB spring) and the other as the SW transformation section (SW spring). Considering a constant mechanical property in both sections, the effective stiffness $E_N$ of DNT with $N$ SW defects can be predicted from

$$\frac{1}{E_N} = \frac{NL_{sw}}{E_{sw}L} + \frac{(L - NL_{sw})}{E_{pb}L} \qquad (3)$$

where $L_{sw}$ and $E_{sw}$ are the effective length and effective local stiffness of the SW defect region, respectively; $E_{pb}$ is the effective local stiffness of the poly-benzene rings region; $N$ and $L$ is the number of SW defect and the sample length, respectively. Fitting Eq. (3) with the MD results with $L_{sw}$, $E_{sw}$, and $E_{pb}$ as fitting parameters, good agreement is found between the spring model and the MD values for DNTs with smaller number of SW defects (Figure 6). Apparently, the length and structure dependent yield strain and strength can be well understood from the serial spring model. However, a vast difference is found for DNTs with larger number of SW defects (see Supporting Information), which signifies the inappropriateness of the linear spring model in describing these DNTs. Such inconsistency originates from the brittle-to-ductile transition when the DNT possesses a relatively large number of SW defects. Thus, to describe the mechanical behaviour of a ductile DNT with the poly-benzene region shorter than approximately 4 nm, the spring model with nonlinear coupling should be adopted.

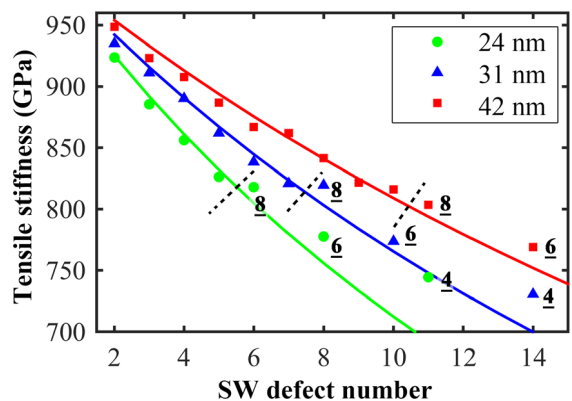

**Figure 6** Comparisons of the tensile stiffness between the spring model and the MD results. The solid lines are fitted using the data from DNT with constituent units longer than DNT-6. Obvious deviation between the spring model and the MD results is observed from DNT-8, and the shorter the poly-benzene (more number of SW defects) the larger the deviation. The underline numbers represent the number of poly-benzene rings between two adjacent SW defects in the DNT structure.

**Conclusions**

In summary, the diamond nanothread (DNT) exhibits intriguing properties that are unseen in other 1D carbon allotropies, such as diamond nanowire and CNT. Besides its excellent mechanical properties, DNTs show a transition from a brittle to a ductile characteristic when the length of its poly-benzene decreases (or the number of SW defect increases), suggesting a tunable mechanical property. Such transition arises from the hardening process of the SW defect under tension. Specifically, the SW defect acts like a grain boundary that interrupts the consistency of the poly-benzene rings in the DNT structure. It is found that the yield strength of the DNT fluctuates in the vicinity of a certain value, and is independent of the "grain size" (i.e., length of poly-benzene) for a given sample length. On the other hand, both yield strength and yield strain show a considerable dependence on the total length, which is due to the fact that the failure of the DNT is dominated by the SW defect. Such intriguing properties of DNT are expected to offer appealing technological applications. Through MD simulations and theoretical analysis, we not only provide insightful understanding on the mechanical properties of the DNT, but also propose the route as a general guide for design of DNT-based device with tunable mechanical properties. Also, the SW defect is expected to endow the DNT with intriguing thermal and electrical properties. As is evidenced from our recent work, the DNT exhibits a superlattice thermal transport characteristics.[41] Moreover, the highly tailorable characteristics of the structure and mechanical properties of DNTs are expected to be found in larger cross-linked systems, such as DNT-based yarn or fiber. Besides the altered individual thread strength due to the cross-links, a similar brittle/ductile transition characteristic is expected for such cross-linked systems. By tuning the distributions and density of the cross-links, the mechanical properties of DNT-based bundles can be controlled.

## AUTHOR INFORMATION

**Corresponding Author**

zhangg@ihpc.a-star.edu.sg; yuantong.gu@qut.edu.au

**Acknowledgement**

Supports from the ARC Discovery Project (DP130102120), the Australian Endeavour Research Fellowship, and the High Performance Computer resources provided by the Queensland University of Technology, and A*STAR Computational Resource Centre (Singapore) are gratefully acknowledged.

**Supporting Information**

Supporting information is available for the discussion on the influence from cut-off distance of the AIREBO potential on DNT, CNT and graphene; the comparison of the mechanical properties between DNTs with evenly and unevenly distributed SW defects; the simulation settings and results for a confined region; and the fitting results using the serial spring model.